\documentclass{optica-article}

\journal{opticajournal}
\articletype{Research Article}

\usepackage{color}
\usepackage{amsmath}

\begin{document}
\title{Bayesian approach to coherent combination of single photon beams}

\author{Antoni Mikos-Nuszkiewicz,\authormark{1,2,3*} Jerzy Paczos,\authormark{3,4} Konrad Banaszek\authormark{2,3} and Marcin Jarzyna,\authormark{1,2}}

\address{\authormark{1}Department of Optics, Palacký University, 17. listopadu 1192/12, 771 46 Olomouc, Czech Republic\\
\authormark{2}Centre for Quantum Optical Technologies, Centre of New Technologies, University of Warsaw, Banacha 2c, 02-097 Warszawa, Poland\\
\authormark{3}Faculty of Physics, University of Warsaw, Pasteura 5, 02-093 Warsaw, Poland\\
\authormark{4}Department of Physics, Stockholm University, SE-106 91 Stockholm, Sweden}

\email{\authormark{*}a.mikos-nuszkiewicz@uw.edu.pl}

\begin{abstract*}
   We theoretically investigate the performance of coherent beam combination of two light beams under relative phase fluctuations in the photon starved regime. We apply a first-principles approach using the optimal Bayesian phase correction protocol. We analyze the efficiency of beam combination as a function of the phase fluctuations strength
\end{abstract*}
\section{Introduction}

Processing and detection of weak optical signals lies at the core of many technological and scientific endeavours, such as ranging \cite{Esteban2011}, imaging \cite{Lemos2014, Kirmani2014, Morris2015} or spectroscopy \cite{Sherlock2009}.  A particular example is deep-space communication, in which the large distance between the transmitter and receiver can limit the received signal power to a level much less than a single photon per time slot \cite{Jarzyna2024}. As a direct consequence, the error rates experienced by the receiver increase and prevent attaining meaningful communication rates. In order to circumvent this issue, it is necessary to increase the received signal power. Unfortunately, the obvious solution, which is to increase the transmit laser power, is often prohibitively difficult due to e.g. mass and electrical power budgets of the spacecraft payload. Alternatively, one may also increase the receiver aperture. However, because of atmospheric turbulence the phase coherence of a beam at two points of a large aperture is lost at distances larger than the Fried parameter \cite{Hardy1998}. As a result, the received signal is divided between multiple spatial modes, each accompanied by noise. One may remedy this issue with the help of adaptive optics that allow one to reduce the number of receiver modes down to one and increasing the signal-to-noise ratio. However, such a technique is expensive to use in combination with large apertures \cite{liuAdaptiveOpticsFreespace2016, stahlSurveyCostModels2010}.

An alternative approach is to utilize the technique of coherent beam combination that gathers light at several small or medium size apertures and then combines it into a single beam in such a way that amplitudes of the input fields are coherently added \cite{Weyrauch2011, Yang2017, larssonCoherentCombiningLowpower2022}. Such a scheme allows one to increase the received power level and, more importantly, leaves the beam for further processing, e.g. detection with sophisticated quantum enhanced measurements \cite{becerraExperimentalDemonstrationReceiver2013, becerraPhotonNumberResolution2015, cookOpticalCoherentState2007}. A major advantage of this approach is its scalability, i.e. one may combine light from as many apertures as required by cascading a basic two-beam scheme.

The concept of coherent beam combination was developed originally for obtaining high-power laser beams which opens up new possibilities in various fields, such as laser material processing \cite{majumdarLaserMaterialProcessing2011}, remote sensing \cite{mehendaleReviewLidarTechnology2020} and laser-induced nuclear fusion \cite{hoganNationalIgnitionFacility2001}. In contrast, the low-power regime became an area of interest only very recently, with fundamental quantum limits investigated theoretically with the help of tools from quantum estimation theory \cite{mullerStandardQuantumLimit2019a} and experimental realizations \cite{larssonCoherentCombiningLowpower2022, hackerPhaselockingInterferometerSinglephoton2023}. Importantly, a number of future space missions are planned to rely on optical communication \cite{biswasNASADeepSpace2019, sodnikDeepspaceOpticalCommunication2017} and developing techniques for efficient signal reception in the photon starved regime may enable higher data rates or considerable savings on telescope infrastructure.

The two main approaches to coherent beam combination are either a digital one or an analog one. In the former method \cite{Geisler2016} the beams are measured separately with a quadrature detector, e.g., homo- or heterodyne measurements. The combined beam is then reconstructed on a computer from the results of individual measurements with means of digital postprocessing. Despite the advantage of relative simplicity of such a scheme, digital methods can be efficiently used only offline and they necessarily require destruction of the light beams in the detection process. Another problem, especially relevant in the weak power regime, is the inevitable shot noise that accompanies any quadrature detection method. Due to these issues in this work we focus therefore on the latter, analog, approach in which one physically combines the beams by maximising the output intensity \cite{Weyrauch2011, Yang2017, Billault2021, larssonCoherentCombiningLowpower2022}. Although such method requires a more sophisticated setup, it does not suffer from the shot noise if direct detection is utilized and, crucially, the output beam can be sent for further optical processing \cite{Yuan2020} which may include sub-shot noise detection methods \cite{Kennedy1973, Cook2007, Guha2011, Becerra2013, Becerra2015}.

In this work, we theoretically investigate coherent beam combination focusing on the single photon regime. In Sec.~\ref{sec:scheme} we describe basic setup for beam combination and relevant noise models. Sec.~\ref{sec:estimation} is devoted to the Bayes phase correction protocol, whereas in Sec.~\ref{sec:simulation} we present results of numerical simulation showing the interplay between the efficiency of beam combination and parameters, such as phase diffusion and initial light intensity. Finally, Sec.~\ref{sec:conclusions} concludes the paper.

\section{Modelling for coherent beam combination}
\label{sec:scheme}

\begin{figure}[t]
\centering
\includegraphics[width=\textwidth]{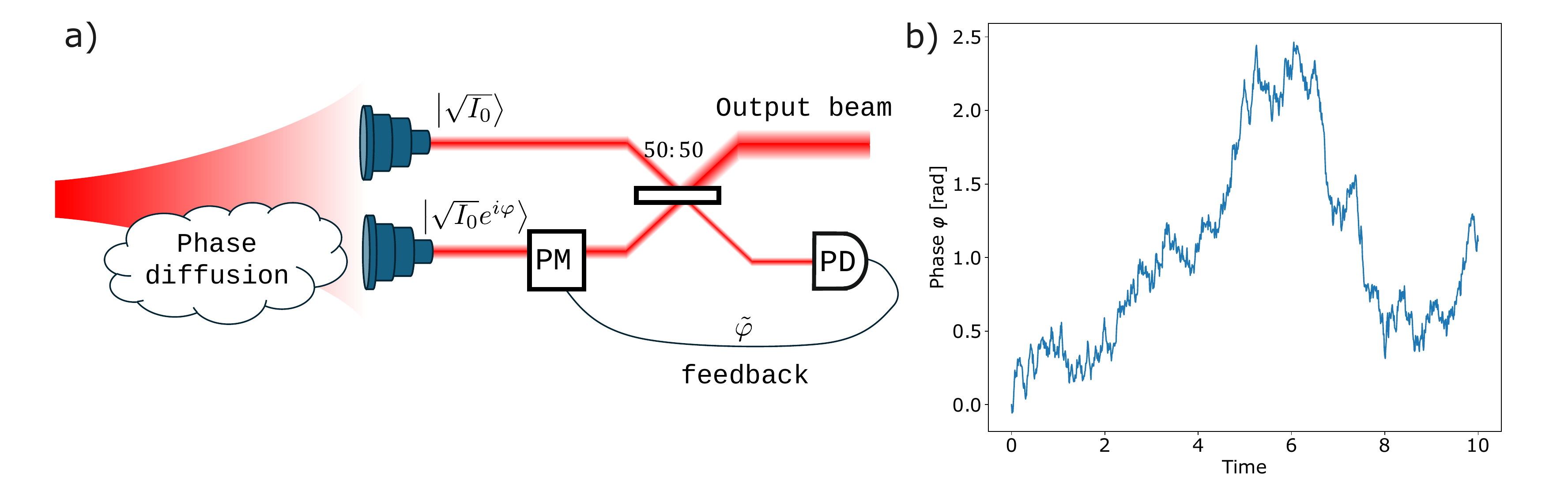}
\caption{a) Scheme for the coherent combination of two beams of light with a relative phase difference $\varphi$ induced by e.g. atmospheric turbulence. Beams gathered by two distinct apertures are interfered on a $50:50$ beam splitter. The dark output port is monitored by a direct photodetector (PD) in the Geiger mode. The measurement results are used to estimate the phase $\varphi$ and compensate it with controlled phase $\tilde{\varphi}$ through a fast feedback loop controlling a phase modulator (PM) on one of the beams. b) Exemplary realisation of phase difference $\varphi$ evolution due to Wiener process in Eq.~(\ref{eq:gauss}) for $D=0.1$ and $\Delta t=0.01$.} \label{Fig:scheme}
\end{figure}

The basic protocol for coherent beam combination is presented schematically in Fig.~\ref{Fig:scheme}(a). The incident light is collected by two individual apertures, resulting in two separate beams. We assume each light beam to be coherent which is the case as long as the apertures do not exceed a certain diameter, however, due to various turbulence effects in the media the light propagates through, their phase is shifted by some unknown value $\varphi$. One of the beams travels through a phase delay $\tilde{\varphi}$ and then they are both interfered on a $50/50$ beam splitter whose dark output port is monitored by a direct detector. The intensities of light leaving the setup through the bright and dark output ports are equal to 
\begin{align}\label{eq:intensity}
    I_{\textrm{bright}}^{\text{out}} &= I_0\left[ 1 + \cos(\varphi - \tilde{\varphi})  \right],\\
    I_{\textrm{dark}}^{\text{out}} &= I_0\left[ 1 - \cos(\varphi - \tilde{\varphi})  \right].
\end{align}
In a standard scenario, based on the intensity measured on the dark port, one can estimate the value of initial relative phase $\varphi$ and, due to a fast feedback loop, tune $\tilde{\varphi}$ such that $I_2^{\textrm{out}}\approx 0$. By keeping this intensity minimal, one ensures constructive interference in the second output port, resulting in all light leaving the setup through there. 

A crucial difference from the standard scheme is that in the low-power regime the measurement of light intensity produces a discrete signal. This is because of the granular nature of light which results in probabilistic detection of individual photons. According to the quantum theory of photodetection \cite{glauberPhotonCorrelations1963, mandelFluctuationsPhotonBeams1958} the number $k$ of photons in a coherent light field measured by a direct detector over an interval $\Delta t$ is described by a Poisson probability distribution
\begin{align}\label{eq:poisson}
p(k) = e^{-\bar{n}}\frac{\bar{n}^k}{k!}, 
\end{align}
where $\bar{n}=\int_{\Delta t} I_D(t) dt$ denotes the average number of photons in a single time slot of duration $\Delta t$ and where $I_D(t)$ is the power in the photon number units. If the intensity does not change significantly during the entire time slot one may approximate $\bar{n}\approx I_D \Delta t$. Since we assume low power, $I_0\Delta t\ll 1$, and $I_D\leq 2I_0$, the only relevant contributions will come from events in which there was only one photon detected or none at all, meaning it is sufficient for the detector to work in the Geiger mode. For such a detector the photodetection probabilities in the $i$-th time interval are explicitly given by
\begin{align}
p_d(k_i=0|\varphi_i) &= \exp\left\{ -I_D\Delta t \left[1 - \cos(\varphi_i - \tilde{\varphi}_i)\right] \right\},\label{Eq:prob_detect1}\\
p_d(k_i=1|\varphi_i) &= 1-p_d(k=0|\varphi_i),\label{Eq:prob_detect2}
\end{align}
where $k_i=0$ denotes no photons measured and, conversely, $k_i=1$ represents a situation in which at least one photon was detected which is predominantly composed of exactly single-photon events. The phases $\varphi_i$ and $\tilde{\varphi}_i$ denote the true relative phase and phase correction in the $i$-th time interval. Note that for higher incident signal powers one would benefit from considering separately events in which more than one photon was detected. On the other hand, depending on the bandwidth of the feedback loop and specific detector characteristics such as dead time, one can reduce the probing time $\Delta t$ and by this decrease the instantaneous signal number of photons to remain in the single photon regime.

\subsection{Relative phase fluctuations}
\label{sec:fluctuations}

In practical scenarios the incident light beams may not be constant in time. Both their relative phase and intensities may vary independently. This is the case for optical signals sent from space that travel through the atmosphere. Even in very good, cloudless observation conditions, the signal experiences scintillations \cite{lyrasScintillationDueAtmospheric}. In order to capture the essential features of the phase correction protocol in our model we assume constant intensities, both equal to $I_0$, and phase fluctuations described by the Wiener process. This means that the relative phase in the $i+1$ -th time interval depends probabilistically on the phase in the $i$-th time interval as
\begin{equation}
    \varphi_{i+1}=\varphi_i+\Delta\varphi_{i+1}, \label{Eq:random_walk}
\end{equation}
where $\Delta\varphi_{i+1}$ is the phase increment described by a Gaussian random variable with distribution
\begin{align}\label{eq:gauss}
    q(\Delta \varphi_{i}) := \mathcal{N}(0, 2D \Delta t) = \frac{1}{\sqrt{4\pi D \Delta t}}\exp{\Big[-\frac{(\Delta \varphi_i)^2}{4 D \Delta t}}\Big],
\end{align}
where the parameter $D$ denotes the phase diffusion per unit time, with corresponding variance $\sigma^2 = 2 D \Delta t$. Note that one has to be careful with this definition as the phase is defined on a $(-\pi,\pi]$ interval, whereas a Gaussian random variable can take any real value. This, however, is not a problem as long as the width of distribution in Eq.~(\ref{eq:gauss}) is small with respect to $2\pi$. A more rigorous approach requires to instead take a distribution describing Gaussian diffusion on a circle \cite{levyProcessusStochastiquesMouvement1965}. An example of the relative phase evolution is shown in Fig.~\ref{Fig:scheme}(b). It is seen that even for small diffusion coefficients the phase may vary considerably on longer time scales.

\section{Bayesian phase correction}
\label{sec:estimation}

A crucial part of any coherent beam combination protocol is the method in which the relative phase between the beams is estimated. As the phase between incident beams changes in time due to fluctuations, one has to perform this estimation in every time step in order to assure continuous constructive interference in the bright output port. For Geiger-type detectors that we consider the only available information in each time step are detector clicks. Intuitively, if no photons were registered in a time step $i$ one should not change the value of the phase correction $\tilde{\varphi}$ since it seems to be close to the true value of the phase difference. On the other hand, when the detector clicks, i.e. a photon is registered, one knows that the phase correction value has to be adjusted. This notion can be formalized by using a Bayes estimator.
\begin{align}
    \tilde{\varphi}_{i+1} = \textrm{argmax}_{\theta\in (-\pi,\pi]}  \big[ p(\theta = \varphi_i | k_i, k_{i-1},\dots,k_1)\big], \label{Eq:estimator}
\end{align}
where $k_i \in \{0,1\}$ denotes whether in the time step $i$ the photon was measured or not. Here, the phase correction value in the $i+1$-th time step is estimated as the maximum of the posterior distribution of phase based on the whole history of detected events up to a time step $i$.

The posterior distribution in Eq.~(\ref{Eq:estimator}) can be calculated from the Bayes theorem.
\begin{align}
p(A|B_i, \dots,B_1) = \frac{p(B_i|A)p(A|B_{i-1},\dots,B_1)}{p(B_i)},
\end{align}
where the events $B_i$ that serve as conditions are independent. Taking $A=\{\theta=\varphi_i\}$ and $B_i=\{k_i\}$ one obtains
\begin{align}
p(\theta = \varphi_i|k_i, \dots,k_1) &= \frac{p_d(k_i|\theta = \varphi_i)p(\theta=\varphi_i|k_{i-1},\dots,k_1)}{p(k_i)}.\label{Eq:Our_bayes}
\end{align}
The probability distribution in the denominator at the right-hand side of the above equation Eq.~(\ref{Eq:Our_bayes}) does not depend on the variable $\theta$. Since one is interested in the maximum over $\theta$ of the whole expression, this term can be disregarded, as it will not change the estimator value. The first term in the product in the nominator is the probability of a photon detection assuming the true relative phase is equal to $\theta$ which can be easily calculated by using Eqs.~\eqref{Eq:prob_detect1}--\eqref{Eq:prob_detect2}. Finally, the second term in the nominator can be expressed in the following way
\begin{align}
    &p(\theta=\varphi_i|k_{i-1},\dots,k_1) =\nonumber\\
    &=p(\theta=\varphi_{i-1}+\Delta \varphi_{i}|k_{i-1},\dots,k_1)=\\
    &=p(\theta=\varphi_{i-1}|k_{i-1},\dots,k_1)*q(\Delta \varphi), \label{Eq:convoluted_distribution}
\end{align}
which is the posterior probability distribution in the previous time step convoluted with a Gaussian distribution describing phase fluctuations. Note that here we made an assumption of weak phase fluctuations, discussed already in Sec.~\ref{sec:fluctuations}, i.e. that the Gaussian distribution $q(\Delta\varphi)$ is narrow. One can therefore safely assume that $\theta\in(\theta_{\max}-\pi,\theta_{\max} +\pi]$, where $\theta_{\max}$ is the maximum of  posterior distribution in the $i-1$-th time step. Combining Eqs. \eqref{Eq:estimator}, \eqref{Eq:Our_bayes} and \eqref{Eq:convoluted_distribution}, one obtains an iterative expression for the posterior distribution in every time step
\begin{align}
    &p(\theta = \varphi_i|k_i, \dots,k_1) = \label{Eq:final_estimator}\\
    &=N_i\cdot p(\theta=\varphi_{i-1}|k_{i-1},\dots,k_1)*q(\Delta \varphi)p_d(k_i|\theta = \varphi_i),\nonumber
\end{align}
where $N_i$ is a normalisation constant. The initial phase distribution is taken to be uniform on the whole interval $(-\pi,\pi]$ as  $p(\theta=\varphi_0)=1/(2\pi)$ since we assume no prior knowledge on the relative phase. Other choices can also be made, potentially improving the speed of convergence of the phase-correction protocol.

\section{Numerical simulations}
\label{sec:simulation}

To investigate the efficiency of the above protocol we perform numerical simulations of beam combining experiment  with on-off direct detection and Gaussian phase diffusion noise. The initial relative phase value $\varphi_0$ was taken at random from the entire range $(-\pi,\pi]$. In every subsequent time step $i$ we draw a new phase value $\varphi_i$ according to Eq.~\eqref{Eq:random_walk}. The photodetection measurement outcome $k_i$ in the dark port is then randomly drawn according to the probabilities from Eqs.~\eqref{Eq:prob_detect1}--\eqref{Eq:prob_detect2}. In each time step $\varphi_i$ is estimated with the help of Bayes' estimator given by Eqs.~\eqref{Eq:estimator}, \eqref{Eq:final_estimator}.The control phase $\tilde{\varphi}_{i+1}$ is subsequently set equal to such obtained estimate. Such procedure guarantees one the highest probability that the control phase equals the true one which would result in perfect beam combination.

\begin{figure}[t]
\centering
\includegraphics[width=1.\textwidth]{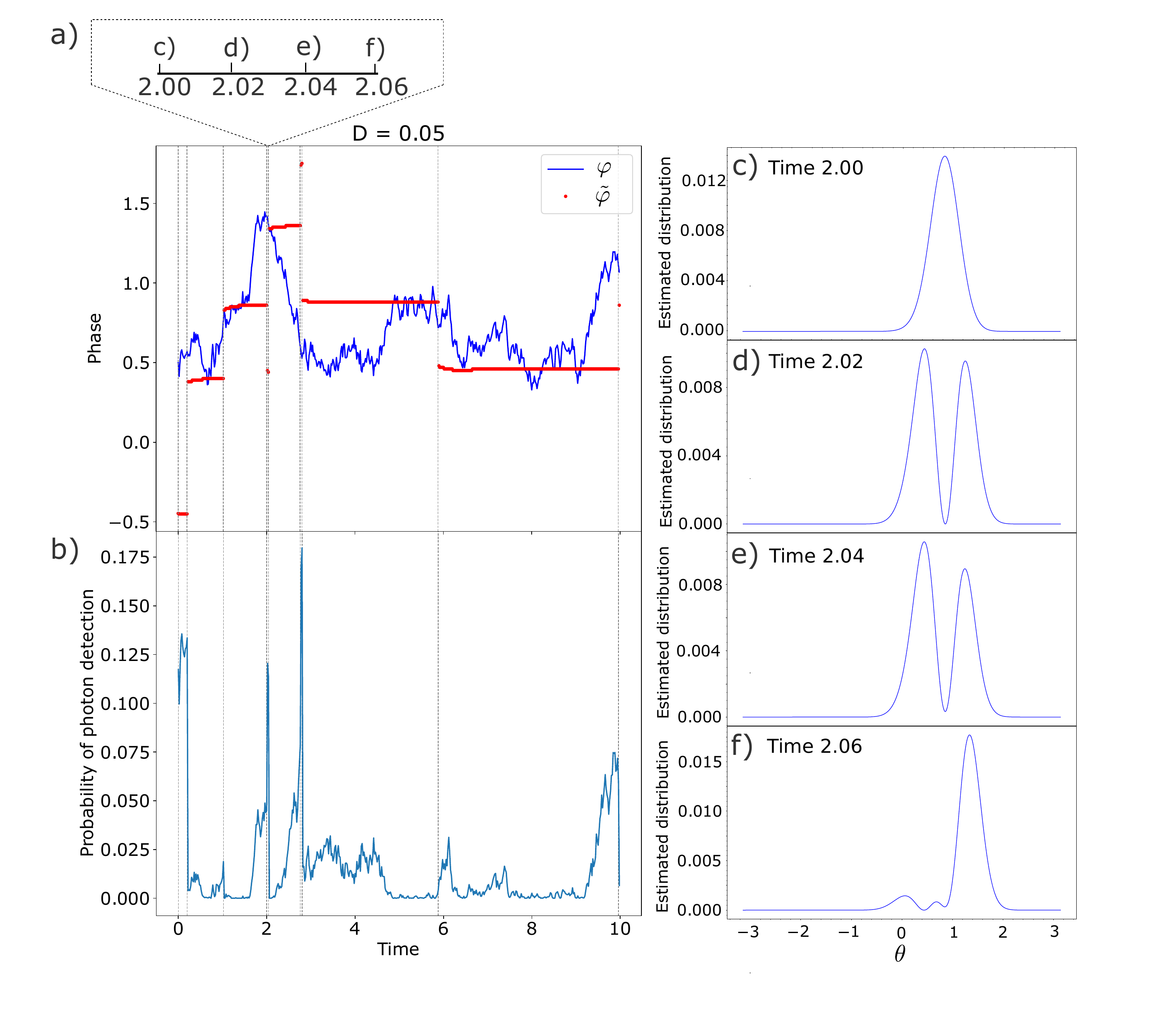}
\caption{(a) Evolution of the true (blue) and control phase (red) and corresponding probability of photodetection (b) obtained from numerical simulation for the diffusion coefficient $D=0.05$, input light intensity $I_0=10$ and time slot duration $\Delta t=0.01$. The dashed vertical lines indicated photodetection events at the dark output port. Exemplary evolution of posterior distribution for the phase in the case of a wrong sign of correction choice at $t=2.0$, starting immediately before photodetection (c), for two subsequent time steps (d-e) followed by a choice of right phase after a second photodetection event.} \label{Fig:movie}
\end{figure}

The evolution of the true and corresponding control phases is presented in Fig.~\ref{Fig:movie}(a). It is seen that the control phase quickly attains a value close to the actual phase in few initial time steps. More importantly, similar behavior can be seen also for later times, i.e. following time steps in which the difference between the two phases is large the control phase quickly changes to approximate the true phase. On the other hand, when the difference between the phases is small, the value of the control phase remains constant. This is a consequence of the fact that the higher the difference $|\varphi_i - \tilde{\varphi}_i|$, the more probable it is to detect a photon in that particular time step, as evidenced in Fig.~\ref{Fig:movie}(b). For small differences, this probability is low and there are no events providing new information about the phase.

An apparent vulnerability of the Bayesian procedure is that the intensity in the dark port Eq.~(\ref{eq:intensity}) and, consequently, the probability distribution in Eqs.~\eqref{Eq:prob_detect1},~\eqref{Eq:prob_detect2} are even functions of the phase difference $\varphi-\tilde{\varphi}$. This means that after every photodetection event the posterior distribution gains an additional local maximum, as  seen in Fig.~\ref{Fig:movie}(d). It may happen therefore that the procedure chooses a wrong maximum ending up with a phase correction value differing considerably from the true phase. An example of such situation can be seen in Fig.~\ref{Fig:movie}(a) for time $t=2.02$. However, such a wrong correction results in a much higher probability of photodetection in consecutive time steps that quickly provide additional phase information. Since the Bayesian estimator is based on the entire history of events, it can therefore correct itself and find the right value of phase correction in several subsequent time steps, as evidenced by Fig.~\ref{Fig:movie}(d-f). A visualization of the estimator behavior is provided in the supplementary material in the form of a short movie.

\begin{figure}[t]
\centering
\includegraphics[width=0.93\textwidth]{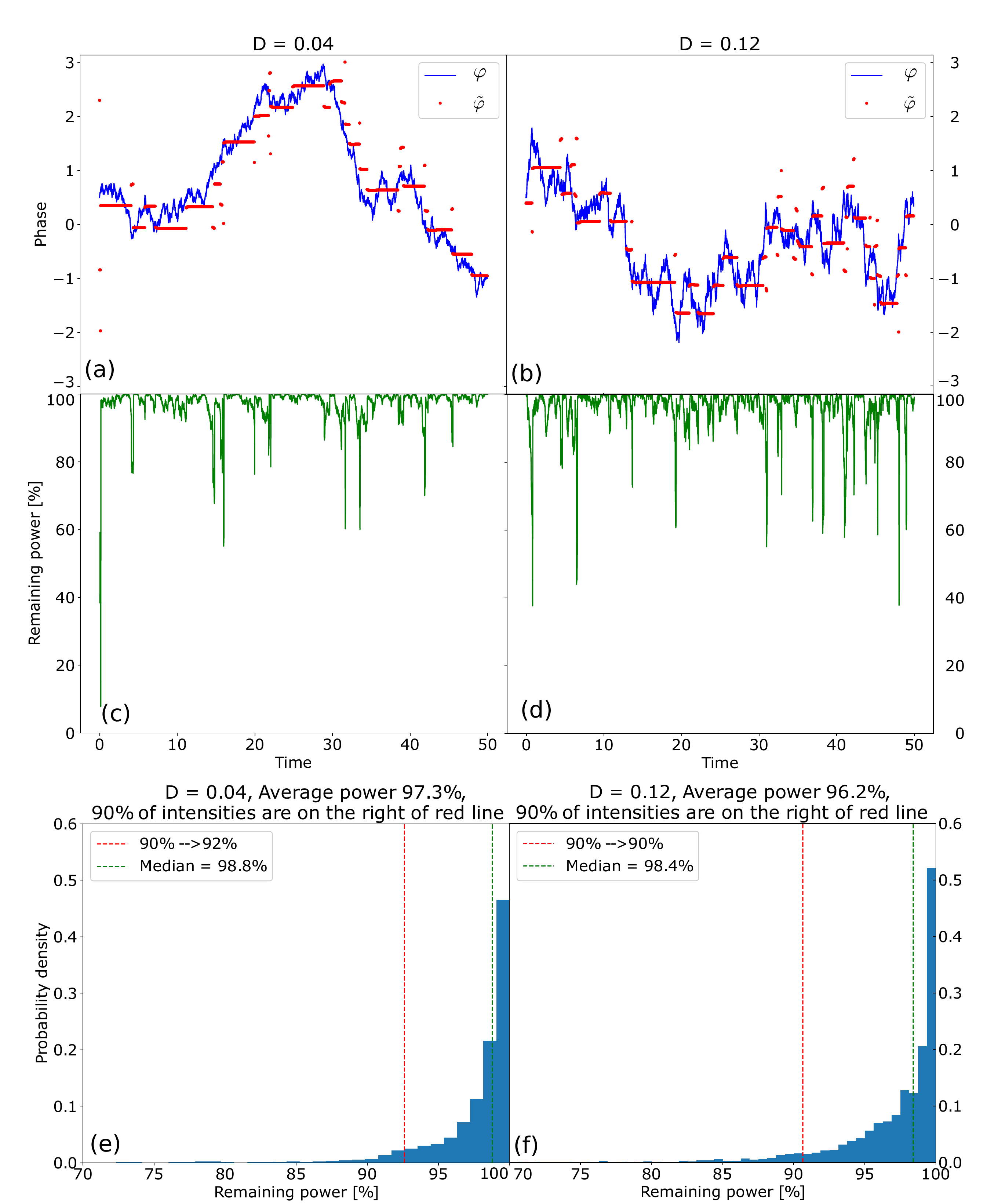}
\caption{Evolution of the true (blue) and control phase (red) obtained from numerical simulation for the input light intensity $I_0=10$, time slot duration $\Delta t=0.01$ and diffusion coefficient $D=0.04$ (a) and $D=0.04$ (b). The dependency of the combined beam's intensity in time, written as a percentage of the incoming intensities (c), (d). (e), (f) the histograms of the intensities from respective plots (c) and (d).} \label{Fig:phase}
\end{figure}

It is seen in Figs.~\ref{Fig:phase}(c), (d) that the light intensity leaving the bright output port of the combination setup is usually close to the maximum value of $2I_0$. The sudden drops are very short and occur when the Bayesian estimator chooses the wrong phase value. As mentioned above, this is quickly corrected by the procedure, as evidenced by the very narrow character of these intensity drops. A more systematic analysis reveals that, depending on the strength of phase noise and input power, the intensity in close to $90\%$ of time steps achieves at least $90\%$ of the maximum possible value, as seen in Fig.~\ref{Fig:phase}(e), (f) for $D\leq 0.12$ and $I_0=10$, $\Delta t=0.01$. The median of the intensity distribution is close to $99\%$.

A relevant quantity to characterize the quality of beam combination in an illustrative way is the beam combination efficiency $\eta$ defined as the ratio of the average intensity leaving the setup through the bright output port over the combined intensities of the input beams $\eta=I_1^{\textrm{out}}/(2I_0)$. Such an average has to be taken over the ensemble of all possible realizations of the stochastic process resulting from the phase diffusion and the correction procedure.  Numerically, one can approximate that value by running the simulation many times for a particular choice of parameters because each simulation ends up in a different realisation. Then, an average intensity for every time step $i$ over all simulations is computed. As a result, one obtains a function of average power in time which should converge to a constant value, since the impact of initial phase difference becomes negligible for later times. However, such an approach is computationally very demanding. We therefore refer to the ergodic theorem \cite{sineReviewUlrichKrengel1988} which states that, provided certain assumptions are satisfied, the average over an ensemble for a stochastic process is equal to its average over time for long times $t\rightarrow \infty$. The intuition behind this approach is that after a sufficiently long time $t$ the true optical phase should take every possible value many times. Therefore, if one possesses a simulated realization from $t_0$ to $t_1$ and $t_0\ll t_1$, and takes an intermediate time $\tau$ such that $t_0 \ll \tau \ll t_1 $, then the part of simulation from $\tau$ to $t_1$ can be considered as a new simulation. This means that taking long total time in one simulation and averaging the results in time is equivalent to taking an average over many independent simulations.

\begin{figure}[t]
\centering
\includegraphics[width=0.85\textwidth]{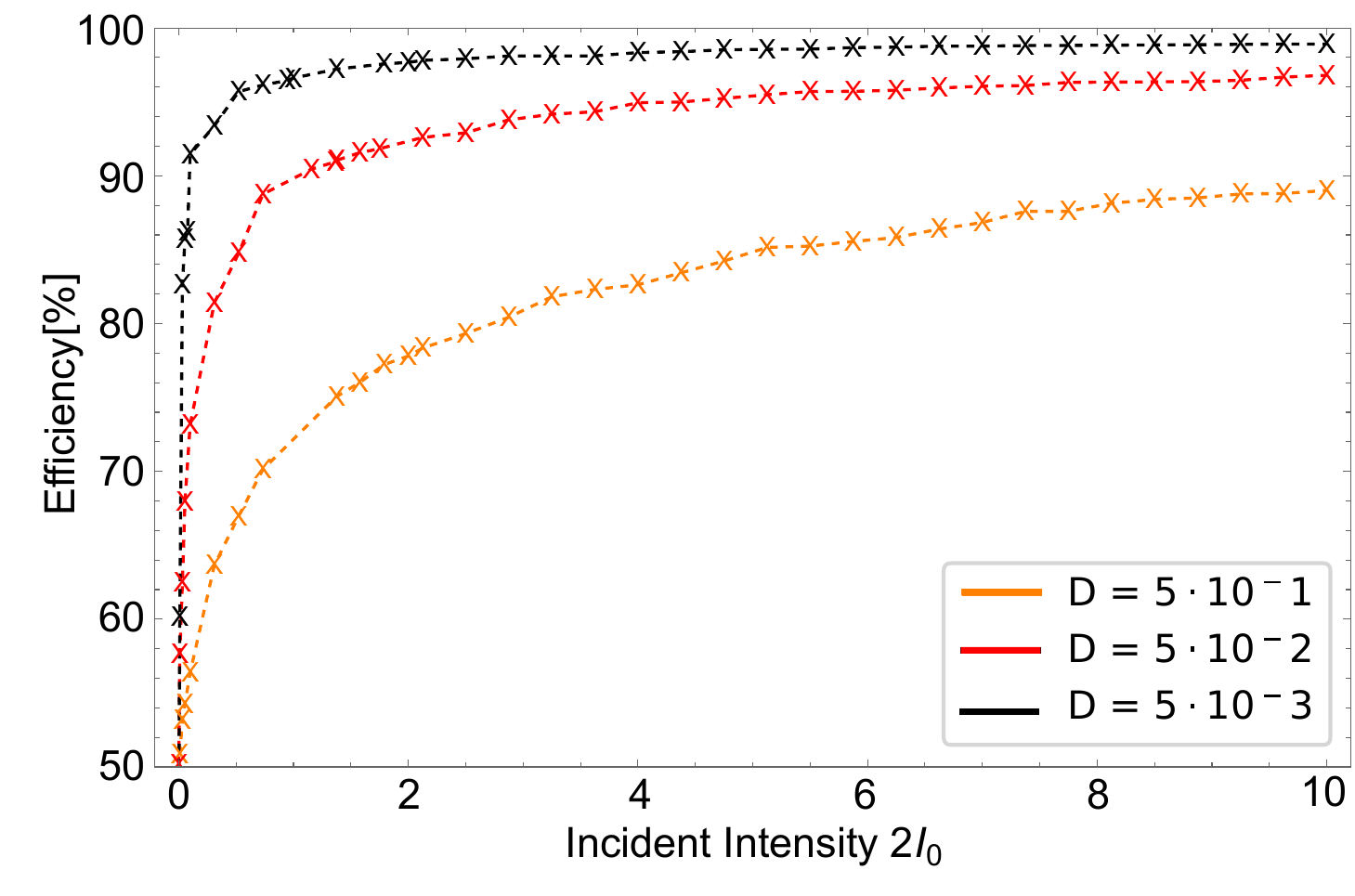}
\caption{Average efficiency of beam combination versus the incident intensity $2I_0$ for three values of a phase diffusion constant $D$ simulated over 50,000 slots. The dashed lines are given as a guide for an eye.} \label{Fig:average}
\end{figure}

The efficiency of the beam combination averaged over time is shown in Fig.~\ref{Fig:average} for three values of the phase diffusion constant $D$. For all values of the diffusion constant it is seen that efficiency increases with incident intensity $I_0$. This is expected since the phase can be corrected only when a photon is detected, and the chance of detection is higher for larger intensities. Due to an increased probability of detection, the procedure is also sensitive for lower phase differences which results in increased stability. A crucial observation from the plot in Fig.~\ref{Fig:average} is that the phase diffusion limits the maximal efficiency that can be attained, even for larger incident intensity values. It also seems that there is a tradeoff  relation between the diffusion constant and incident intensity below which the beam combination efficiency drops considerably. This is because for very weak intensity the phase may differ significantly between two photodetection events, meaning the information gathered by the phase correction protocol in the preceding time steps is insufficient. The protocol then effectively works as if after every detector click the phase was completely random, as in the beginning of the procedure. In other words, the phase diffusion sets up a characteristic timescale in which one has to register at least one photon in order to correct the phase. If the intensity is too low, the probability of such event becomes small and beam combination stops being possible.

\section{Conclusions} \label{sec:conclusions}
We considered the problem of coherent beam combination for weak signals in the photon-starved regime with direct detectors operating in the Geiger mode. We used Bayesian phase correction protocol under Gaussian phase noise and constant intensity assumption. It has been found that one can efficiently combine the beams up to low intensities, depending on the strength of phase diffusion. However, for very small intensities beam combination stops being effective due to low probability of photodetection in a timescale set up by the diffusion constant. Importantly, since Bayesian phase estimator takes into account the whole history of events, it is optimal, which means that this limit is fundamental and is caused by the discrete nature of photocounts. Note that our results apply also to stronger beams, as for efficient combination procedures, at some point, the intensity incident at the dark port becomes small. Our results suggest that coherent beam combination is feasible in many applications characterized by low signal level, in particular in deep-space optical communication and quantum optical communication.

In our work we assumed that the bandwidth of the feedback loop employed in the phase correction is much greater than the characteristic timescale of phase fluctuations. If that is not the case, the performance should drop the longer it takes to correct the phase, meaning one requires larger incident intensity. This may be a predominant limiting factor in instances in which phase fluctuations are large. Another important source of imperfections are also fluctuations of the incident intensity value, which may mimic the choice of correct phase in case they decrease the intensity. A remedy may be to divert some small part of light for an intensity measurement, which would provide additional information for phase correction. A similar effect that, unfortunately, cannot be easily remedied is caused by the additive noise, which prevents the photodetection probability from going down to low values, even when the correction is very close to the true phase value. The exact impact of these effects on beam combination efficiency for weak signals require further investigation.

%
\bibliography{beamComb}
\end{document}